\documentclass[12pt]{iopart}
\usepackage{iopams}
\usepackage{graphicx}
\usepackage{dcolumn}
\usepackage{bm}
\usepackage{amssymb}
\usepackage{verbatim}
\usepackage{epstopdf}
\usepackage{cite}

\newcommand{\abs}[1]{\ensuremath{\left| #1 \right|}}

\newcommand{\beq}{\begin{equation}}
\newcommand{\eeq}{\end{equation}}
\newcommand{\bea}{\begin{eqnarray}}
\newcommand{\eea}{\end{eqnarray}}

\newcommand{\commentout}[1]{{}}

\newcommand{\eva}[1]{\left<#1\right>}

\newcommand{\threevec}[3]{\left(\begin{array}{c}#1\\#2\\#3\end{array}\right)}
\newcommand{\nematic}{\mathbf{\hat{d}}}

\newcommand{\absF}{|\langle\mathbf{\hat{F}}\rangle|}
\newcommand{\expF}{\langle\mathbf{\hat{F}}\rangle}
\newcommand{\SO}{\mathrm{SO}}
\newcommand{\U}{\mathrm{U}}

\newcommand{\rr}{\ensuremath{\mathbf{r}}}

\newcommand{\rhat}{\ensuremath{\mathbf{\hat{r}}}}
\newcommand{\xhat}{\ensuremath{\mathbf{\hat{x}}}}
\newcommand{\yhat}{\ensuremath{\mathbf{\hat{y}}}}
\newcommand{\zhat}{\ensuremath{\mathbf{\hat{z}}}}

\begin{document}
\title[Imprinting a topological interface using Zeeman shifts]{Imprinting a topological interface using Zeeman shifts in an
  atomic spinor Bose-Einstein condensate}
\author{M O Borgh, J Lovegrove and J Ruostekoski}
\address{Mathematical Sciences, University of Southampton,
Southampton, SO17 1BJ, UK}

\begin{abstract}
We propose to use spatial control of the Zeeman energy shifts
in an ultracold atomic gas to engineer an interface between
topologically distinct regions. This provides an experimentally
accessible means for studying the interface physics of topological
defects and textures.  Using the spin-1 Bose-Einstein condensate as an
example, we find spinor wave functions that represent defects and
textures continuously connecting across the interface between polar
and ferromagnetic regions induced by spatially varying Zeeman shifts.
By numerical energy-minimization we characterize the defect core
structures and determine the energetic stability. The techniques proposed
could potentially be used in the laboratory to emulate
complex interface physics arising, e.g., in cosmological and
condensed-matter contexts in both uniform and lattice systems.
\end{abstract}


\maketitle

\section{Introduction}

The physics of topological defects, such as vortices, becomes
especially intriguing
at the interface between coexisting, topologically distinct phases of a
macroscopically coherent system.
Due to different broken symmetries on either side, a
defect cannot perforate the interface unchanged. Instead it must
either terminate, or continuously connect across the boundary to an
object representing a
different topology.  This situation arises, for example, at the
interface between the $A$ and $B$ phases of superfluid liquid
$^3$He~\cite{salomaa_nature_1987,volovik,finne_rpp_2006}, at
interfaces between regions of different vacua in theories of the early
universe~\cite{kibble_jpa_1976,vilenkin-shellard}, in the physics
of branes in superstring
theory~\cite{dvali_plb_1999,sarangi_plb_2002}, and in exotic
superconductivity~\cite{bert_nphys_2011}.

The parallels~\cite{volovik} between cosmological objects and defects in
superfluids prompted the suggestion that analogues of cosmological
phenomena can be studied in the
laboratory~\cite{volovik,zurek_nature_1985}, for example the formation of
defects in phase
transitions~\cite{bauerle_nature_1996,ruutu_nature_1996,weiler_nature_2008,sadler_nature_2006} or
properties of cosmic vortons (superconducting cosmic strings~\cite{Radu})~\cite{ruostekoski_prl_2001,battye_prl_2002,savage_prl_2003,ruostekoski_pra_2004,kawakami_prl_2012}.
Current experimental techniques in atomic physics allow accurate
measurements and precise
control and manipulation of ultracold atomic gases by finely tuning
electromagnetic fields.
In spinor
Bose-Einstein condensates (BECs), where the atoms retain their spin
degree of freedom, experiments have demonstrated controlled
preparation of coreless
vortices and analogous non-singular
textures formed by the nematic axis~\cite{leanhardt_prl_2003,leslie_prl_2009,choi_prl_2012,choi_njp_2012}.
Vortex nucleation in phase
transitions~\cite{sadler_nature_2006} and dynamical
formation of spin
textures~\cite{vengalattore_prl_2008,kronjager_prl_2010,bookjans_prl_2011,guzman_pra_2011}
have also been experimentally observed.
Simultaneously there has been a rapidly
increasing theoretical interest in the wide variety of vortices, point
defects and particle-like textures in
two-component
(pseudospin-$1/2$)~\cite{ruostekoski_prl_2001,al-khawaja_nature_2001,mueller_prl_2002,battye_prl_2002,savage_prl_2003,kasamatsu_prl_2003,ruostekoski_pra_2004,mason_pra_2011,kawakami_prl_2012}, 
as well as
spin-1~\cite{ho_prl_1998,ohmi_jpsj_1998,yip_prl_1999,leonhardt_jetplett_2000,mizushima_prl_2002,martikainen_pra_2002,zhou_ijmpb_2003,stoof_monopoles_2001,ruostekoski_monopole_2003,savage_dirac_2003,pietila_prl_2009_dirac,reijnders_pra_2004,mueller_pra_2004,saito_prl_2006,takahashi_pra_2009,ruokokoski_pra_2011,simula_jpsj_2011,kobayashi_pra_2012,lovegrove_pra_2012,lovegrove_prl_2014}
and
spin-$2,3$~\cite{semenoff_prl_2007,huhtamaki_pra_2009,kobayashi_prl_2009,santos_spin-3_2006,barnett_pra_2007}
BECs.
This development brings multi-component systems of ultracold atoms
to the forefront as candidate laboratories where properties of a
variety of field-theoretical
solitons (see for exampel~\cite{Radu,bogomolny_sjnp_1976,jackiw_prd_1976,manton-sutcliffe,faddeev_nature_1997})
may be studied.

We have previously suggested~\cite{borgh_prl_2012,borgh_pra_2013} that
spatially non-uniform manipulation of scattering lengths
by optical or microwave-induced Feshbach resonances can be used to
study the physics of topological interfaces in ultracold atomic gases with spin
degree of freedom. An example is the spinor BECs, which exhibit
distinct phases of the ground-state
manifold. In the simplest case of a spin-1 BEC there are two phases,
polar and FM,
and the sign of the spin-dependent interaction determines which phase
is energetically favourable.
We proposed that a combination of (microwave or optical) Feshbach
resonances and spatially-dependent AC-Stark shifts can be used to enforce
different signs of this interaction in different spatial regions of
the same spin-1 BEC, establishing a coherent interface between the
phases.  Within this system, we formulated spinor wave functions
corresponding to defect combinations that can be phase imprinted using
existing techniques.  By numerical simulation we found examples of
energetically stable interface-crossing defects and complex
core deformations, such as the formation of an arch-shaped
half-quantum vortex on the interface. In addition,
defects at an energetically established boundary in a two-component
BEC, where in one region the two components are miscible and in the
other immiscible, have recently been studied in
BECs in~\cite{kasamatsu_jhep_2010,takeuch_jltp_2011,nitta_pra_2012,takeuchi_prl_2012,kasamatsu_jphyscm_2013}.

Here we propose to employ precise spatial engineering of the Zeeman
shifts to create topologically dissimilar regions within a spinor BEC,
providing an experimentally simple route for studying defects and
textures at the emerging topological interface. The ground state of the
spinor BEC generally depends on the linear and quadratic energy shifts
of the Zeeman levels.
In the case of the spin-1 BEC, the Zeeman shift
can cause the condensate to adopt the FM phase even when the polar
phase is favoured by the interactions, and vice
versa~\cite{stenger_nature_1998,koashi_prl_2000,zhang_njp_2003,murata_pra_2007,sadler_nature_2006,ruostekoski_pra_2007}.

In particular, we demonstrate that a stable, coherent, topologically
non-trivial interface between
FM and polar phases of a spin-1 BEC can be
established through spatially non-uniform linear or quadratic Zeeman
shifts.
Uniform ground-state solutions exist, for both polar and FM interaction
regimes, that follow the variation of the
Zeeman shift; the corresponding wave functions continuously
interpolate between the polar and FM
phases. We then analytically construct defect states that continuously
connect defects and textures representing the topology of the FM and
polar phases, such that the connection is provided directly by the
spatial dependence of the Zeeman energy shifts. We show that the
modulation of the Zeeman splitting allows the preparation of a rich family
of interface-crossing defect solutions, with various combinations
of singular (integer and half-quantum) and non-singular vortices,
point defects, and terminating vortices. 
By numerical simulation, we determine the stability properties of the
constructed solutions and determine their energy-minimizing core
structures.

In the polar interaction regime, the interface is established by a
varying linear Zeeman shift. Examples of energetically stable
interface-perforating 
defect configurations in a rotating trap include a singly quantized FM
vortex line continuously 
connecting to a singly quantized polar vortex whose core splits into a
pair of half-quantum vortices, 
as well as a polar vortex that terminates at the interface.

For a BEC in the FM interaction regime, a spatially varying quadratic
energy shift is used to establish the interface between the polar and FM
phases. We find an energetically stable, singular FM vortex that
terminates at the interface. Moreover, non-singular, coreless vortices 
in the FM phase become energetically favourable, and we find
energetically stable structures 
where the coreless vortex continuously connects to a singly quantized
vortex on the polar side of the interface. The unusual property of the
singly quantized polar vortex in this configuration is the axially
symmetric stable vortex core, in which the line singularity is
filled with atoms in the FM phase, and the core is not split into a
pair of half-quantum vortices. 

The existence of stable core structures of different symmetries in
atomic spinor BECs is reminiscent of the rich vortex core symmetries
encountered in superfluid liquid $^3$He~\cite{salomaa_rmp_1987}. For
example, the core of a singular $B$-phase vortex may analogously
retain a non-zero 
superfluid density by filling with the $A$
phase, either with an axially symmetric core~\cite{salomaa_prl_1983}
or by breaking the axial symmetry when forming a two-fold symmetric
split core~\cite{salomaa_prl_1986,thuneberg_prl_1986}. 

In the case of both FM and polar interactions we also find stable core
structures of singular FM vortex lines terminating on a point defect
in the polar phase. Such a point defect is analogous to the
't~Hooft-Polyakov
monopole~\cite{t-hooft_npb_1974,polyakov_jetplett_1974} and the
combined defect configuration of
the terminating vortex line and the point defect is closely related to
boojums that can exist in superfluid liquid
$^3$He~\cite{vollhardt-wolfle,volovik,mermin_qfs_1977}.
The core of the point defect minimizes its energy by deforming into a
half-quantum line defect connecting at both ends to the interface.
As the point defect, or the `Alice arch' line defect, does not
couple to the trap
rotation, the defect experiences a trivial instability with respect to
drifting out of the atom cloud as a result of the density gradient of
the harmonic trap (the order parameter bending energy of defects and
textures generally favours lower atom densities), but can be otherwise
stable. Such an instability 
could be overcome by creating a local density minimum close to the
trap centre by an additional optical
potential~\cite{ruostekoski_monopole_2003}.

The interface physics with the Zeeman shifts provides several
promising experimental scenarios. 
Accurate tuning of Zeeman shifts has been experimentally demonstrated in
ultracold atoms~\cite{gerbier_pra_2006}, and also applied to the
study of spin textures~\cite{guzman_pra_2011}. On the other hand, the
control of multiple 
interfaces and their time-dependence could
open up avenues for emulating complex cosmological phenomena in the laboratory.
For instance, in superfluid liquid $^3$He~\cite{bradley_nphys_2008}
or in a two-component BEC
system~\cite{kasamatsu_jhep_2010,takeuch_jltp_2011,nitta_pra_2012,takeuchi_prl_2012}
 it has been proposed that
colliding interfaces or phase boundaries could mimic cosmic defect formation.
In a spin-1 BEC we could envisage, for instance, the following set-up:
A disc of polar phase is created in an otherwise FM condensate by locally
increasing Zeeman shift. The two parallel FM-polar interfaces can then be
interpreted as analogues of string-theoretical $D$-branes and
anti-branes arising in theories of brane
inflation~\cite{dvali_plb_1999}. Removing the
local Zeeman shift causes the interfaces to collapse, simulating
defect formation in 
brane annihilation scenarios.  Similar experiments have been performed
with colliding superfluid $^3$He $A$-$B$
interfaces~\cite{bradley_nphys_2008}, where, however, observation of
defects is more difficult.

\section{Effects of Zeeman energy shifts in the spin-1 BEC}

Here we consider the engineering of a topological interface by
manipulation of Zeeman shifts in the context of a spin-1 BEC.
In the Gross-Pitaevskii mean-field theory, the condensate
wave function is a three-component vector $\Psi=\sqrt{n}\zeta$,
where $n$ is the atomic density and $\zeta$ is a normalized spinor
($\zeta^\dagger\zeta=1$) in
the basis of spin projection onto the $z$ axis.
A magnetic field in
the $z$ direction leads to linear and quadratic energy shifts of the
Zeeman sublevels, of strengths $p$ and $q$ respectively.
The Hamiltonian density may then be written as~\cite{kawaguchi_physrep_2012}
\begin{equation}
  \label{eq:hamiltonian-density}
    \fl {\cal H} =  \frac{\hbar^2}{2m}\abs{\nabla\Psi}^2 + V(\rr)n
    + \frac{c_0}{2}n^2
    + \frac{c_2}{2}n^2\abs{\mathbf{\eva{\hat{F}}}}^2
    - pn\eva{\hat{F}_z}
    + qn\eva{\hat{F}_z^2}\,,
\end{equation}
where $V(\rr)$ is the external trapping potential for the atoms.  The
local spin vector is given by the expectation value of the spin
operator $\mathbf{\hat{F}}$ defined as a vector of spin-1 Pauli matrices.
The
contact interaction between the atoms separates into spin-independent
and spin-dependent contributions.  The respective interaction
strengths are
$c_0=4\pi\hbar^2(2a_2+a_0)/3m$ and $c_2=4\pi\hbar^2(a_2-a_0)/3m$,
where $m$ is the atomic mass, and $a_{0,2}$ are the scattering lengths in
the spin-$0,2$ channels of colliding spin-1 atoms.  The
interaction terms give rise to the density and spin healing lengths
\begin{equation}
\label{eq:xi}
\xi_n = \frac{\hbar}{\sqrt{2mc_0n}}, \qquad
\xi_F = \frac{\hbar}{\sqrt{2m|c_2|n}},
\end{equation}
that describe the length scales over which perturbations of the atom
density and the spin magnitudes, respectively, heal.

When the Zeeman shifts are not present ($p=q=0$),
(\ref{eq:hamiltonian-density}) is invariant under spin rotations.
The ground state of the uniform system ($V(\rr)=0$) then exhibits two
phases depending on the
sign of $c_2$. In the FM phase, favoured when $c_2<0$ (e.g., in
$^{87}$Rb), the spin is maximized: $\absF=1$ for a uniform spin
texture. All physically
distinguishable, degenerate, ground states are then coupled by
three-dimensional spin rotations.  The family of FM spinors can
therefore be parametrized as~\cite{ho_prl_1998}
\begin{equation}
  \label{eq:ferro}
  \zeta^{\rm f} 
  = \frac{e^{i\phi^\prime}}{\sqrt{2}}
    \threevec{\sqrt{2}e^{-i\alpha}\cos^2\frac{\beta}{2}}
             {\sin\beta}
             {\sqrt{2}e^{i\alpha}\sin^2\frac{\beta}{2}},
\end{equation}
where $(\alpha,\beta,\phi^\prime)$ are Euler angles defining the spin
rotation such that $\expF =
\cos\alpha\sin\beta\xhat+\sin\alpha\sin\beta\yhat+\cos\beta\zhat$. A
condensate phase $\phi$ is absorbed by the third Euler
angle $\gamma$ to form
$\phi^\prime=\phi-\gamma$, and corresponds
to spin rotations about the local spin direction.
The order-parameter manifold, the broken symmetry in the ground state, is
therefore $\SO(3)$, which supports only two distinct classes of line
defects: singular, singly quantized vortices, and non-singular
coreless vortices (see \ref{sec:appendix}).

The polar phase with minimized spin, $\absF=0$ in the uniform texture, is
favoured when $c_2>0$ (e.g., in $^{23}$Na).
The degenerate ground
states are then characterized by a macroscopic condensate phase $\phi$
and a unit vector
$\nematic$~\cite{leonhardt_jetplett_2000,ruostekoski_monopole_2003}:
\begin{equation}
\label{eq:nematic}
  \zeta^{\rm p} = \frac{e^{i\phi}}{\sqrt{2}}
                 \threevec{-d_x+id_y}{\sqrt{2}d_z}{d_x+id_y}.
\end{equation}
Note that $\zeta(\phi,\nematic)=\zeta(\phi+\pi,-\nematic)$.  These
states are therefore identified, and hence $\nematic$ should be understood as
unoriented.  The identification is reflected in the factorization by
the two-element group in the corresponding broken ground-state symmetry
$[S^2\times\U(1)]/\mathbb{Z}_2$.  This so-called nematic order leads to the
existence of half-quantum vortices [e.g.,~(\ref{eq:hq})]. While all
circulation-carrying vortices are singular in the polar phase, it is
possible to form a non-singular \emph{nematic coreless
  vortex}~\cite{lovegrove_prl_2014},
characterized by a fountain-like texture in $\nematic$
[e.g.,~(\ref{eq:nematic-coreless})]. 

Here we consider the case when either or both of the Zeeman energy
contributions are non-zero.
The linear Zeeman shift in a magnetic field $\mathbf{B}=B\zhat$ is
given by $p=-g_F\mu_BB$, where the
Land\'{e} factor $g_F=-1/2$ in the $F=1$ ground-state manifold of
$^{23}$Na or $^{87}$Rb.  The linear shift can be given a spatial
dependence by careful engineering of the applied magnetic field
$\mathbf{B}$. In alkali-metal atoms in the regime relevant to our
considerations, the quadratic shift $q$, which can be
obtained from the Breit-Rabi formula~\cite{corney}, is positive and
smaller than $p$.  However, by combining a static magnetic field with
an off-resonant  microwave dressing field, accurate tuning of the
quadratic energy shift can be achieved through the
resulting AC-Stark shifts~\cite{gerbier_pra_2006}, or could be induced
by lasers~\cite{santos_pra_2007}.  

When the Zeeman shifts are included, the coupled Gross-Pitaevskii
equations for the spinor components $\psi_j=n\zeta_i$ ($j=+,0,-$)
derived from (\ref{eq:hamiltonian-density}) read
\begin{equation}
  \label{eq:gpe}
  i\hbar\frac{\partial}{\partial t}\Psi =
  \left[-\frac{\hbar^2}{2m}\nabla^2 + V(\rr)
  + c_0n + c_2n\expF\cdot\mathbf{\hat{F}}
  - p\hat{F}_z + q\hat{F}_z^2\right]\Psi.
\end{equation}
In a uniform system, these may be solved
analytically~\cite{zhang_njp_2003,murata_pra_2007,ruostekoski_pra_2007}
to find the stationary states. The Zeeman shifts break the
spin-rotational symmetry of 
the FM and polar ground states. One then finds, in addition to the
purely FM state with
$\expF=\pm\zhat$ and the polar state with $\nematic=\zhat$,
also the steady-state solution~\cite{zhang_njp_2003,ruostekoski_pra_2007}
\begin{equation}
  \label{eq:zetaP}
  \zeta = \frac{1}{\sqrt{2}}\threevec{e^{i\chi_+}P_+}{0}{e^{i\chi_-}P_-},
\end{equation}
where $P_{\pm}=\sqrt{1\pm p/c_2n}$.
The solution (\ref{eq:zetaP}) is valid provided
that the linear Zeeman shift is sufficiently small, such that
$\abs{p}\le c_2n$.
Note that the expectation value of the spin is no longer zero,
$\expF=p/(c_2n)\zhat$, and $\nematic$ lies in the $xy$ plane.  For
very weak linear Zeeman shift $p$, the expression then approaches the polar
state $\zeta = (e^{i\chi_+}/\sqrt{2},0,e^{i\chi_-}/\sqrt{2})^T$.  At
the limit of validity, on the other hand, it coincides with
the FM solution $\zeta = (e^{i\chi_+},0,0)^T$ for $p>0$
[$\zeta = (0,0,e^{i\chi_-})^T$ for $p<0$].
The spinor~(\ref{eq:zetaP}) also represents the lowest-energy state
when $c_2>0$ 
and $q \leq p^2/2c_2n$~\cite{zhang_njp_2003,ruostekoski_pra_2007}. Hence
in a condensate with polar interactions, such as for $^{23}$Na,
(\ref{eq:zetaP}) provides an energetically stable solution that takes values
between FM and polar phases, depending on the linear Zeeman
shift.

\begin{figure}[hbt]
  \begin{center}
    \includegraphics[width=0.8\textwidth]{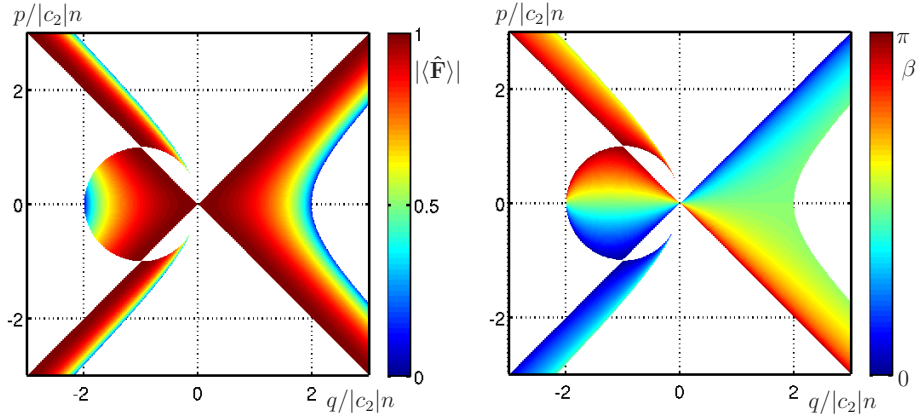}
  \end{center}
  \caption{
    The solution~(\ref{eq:qsol}) is valid in the coloured regions. Left:
    spin magnitude showing interpolation between FM and polar
    limits. Right: angle $\beta$ between $\expF$ and $\zhat$. The figure shows
    the case $c_2<0$ where~(\ref{eq:qsol}) forms the ground state for
    $q>0$. For $c_2>0$, the figures are mirrored around the origin, and
    the solution is never the ground state.}
  \label{fig:q-validity}
\end{figure}

A further solution with variable $\absF$ is given by the
FM-like
spinor~\cite{murata_pra_2007,ruostekoski_pra_2007}
\begin{eqnarray}
  \label{eq:qsol}
  \zeta_\pm = e^{i(\chi_0\mp\chi_z)}(q\pm p)
                \sqrt{\frac{-p^2+q^2+2c_2nq}{8c_2nq^3}}, \\\nonumber
  \zeta_0 =
                e^{i\chi_0}\sqrt{\frac{(q^2-p^2)(-p^2-q^2+2c_2nq)}{4c_2nq^3}}.
\end{eqnarray}
The solution is valid when the expressions under the square roots are
positive.  The corresponding regions in the $(p,q)$ plane are shown in
figure~\ref{fig:q-validity}.  While several regions of validity exist
for both signs of $c_2$, we note that~(\ref{eq:qsol}) forms the ground
state in the uniform system only for $c_2<0$ in the region
defined by $|q|>|p|$ and $p^2>q^2-2|c_2|nq$.  From this point on, we will
consider the solution (\ref{eq:qsol}) only in this parameter
range. The spin vector is in
general tilted with respect to the magnetic field and for
$\chi_0=\chi_z=0$ lies in the $xz$ plane for the parameters of
interest. Then
\begin{equation}
  \label{eq:q-spin0}
  \expF = \frac{\sqrt{\left(q^2-p^2\right)
                \left[\left(p^2-2c_2nq\right)^2-q^4\right]}}
	       {2\abs{c_2}nq^2} \xhat
	+ \frac{p\left(-p^2+q^2+2qc_2n\right)}{2c_2nq^2} \zhat,
\end{equation}
such that
\begin{equation}
  \absF =
  \sqrt{\frac{2q^2\left(p^2+2c_2^2n^2\right)-q^4-p^4}{4q^2c_2^2n^2}}.
\end{equation}
Assuming $p>0$ ($p<0$ analogous by symmetry), the limit ($p=q$) yields
$\absF=1$, corresponding to the FM state
$\zeta=(1,0,0)^T$.  Similarly, $p^2=q^2-2|c_2|nq$ yields the polar limit
$\zeta=(0,1,0)^T$ with $\absF=0$ and $\nematic=\zhat$. From these
results it follows (see
also figure~\ref{fig:q-validity}) that varying $p$ and/or $q$
can continuously connect the two limits while simultaneously rotating
the spin vector from $\expF=\zhat$ in the FM limit to the $x$
direction (implying a
simultaneous rotation of $\nematic$ from $-\xhat$ to $\zhat$). For
simplicity, we will here only consider variations of $q$ for constant $p$.

Together~(\ref{eq:zetaP}) and~(\ref{eq:qsol}) thus provide us with
spinor wave functions that---in the polar and FM interaction regimes,
respectively---represent solutions of different spin magnitudes $\absF$,
depending on the value of the Zeeman shifts.
Here we propose to create the topological interface between FM and
polar phases by
engineering the spatial dependence of the linear and quadratic level
shifts. For spatially varying
Zeeman energy shifts the solutions~(\ref{eq:zetaP})
and~(\ref{eq:qsol}) continuously
interpolate between the polar and FM phases. We will show that a
stable, coherent
interface forms in the intermediate region.  It then
becomes possible for topological defects and textures in the two
regions of the polar and FM phases to connect continuously across the
interface. 
We now proceed to explicitly construct such analytic defect solutions for the
two interaction regimes separately.

\section{Interface by linear Zeeman shift for $c_2>0$}
\label{sec:p-analytics}

In order to construct defect states
in the BEC with polar interactions it is beneficial to transform the
ground-state 
solution (\ref{eq:zetaP}) by applying a BEC phase $\phi$ and a
rotation [determined by the Euler angles $(\alpha, \beta, \gamma)$] of
the 
orthogonal vector triad $(\expF,\nematic,\expF\times\nematic)$. We obtain
\begin{equation}
\zeta=\frac{e^{i\phi}}{2}
      \threevec{\sqrt{2} e^{-i \alpha}\left(e^{ i \gamma}
                \sin^2\frac{\beta}{2}P_-
		-e^{-i \gamma}\cos^2\frac{\beta}{2}P_+\right)}
	       {-\sin\beta\left(e^{i \gamma} P_-+e^{-i \gamma} P_+\right)}
	       {\sqrt{2}e^{i \alpha}\left(e^{i  \gamma}
		 \cos^2\frac{\beta}{2}P_-
		 - e^{-i \gamma}\sin^2\frac{\beta}{2}P_+\right)},
\label{eq:p-general}
\end{equation}
where we have set $\chi_+=\pi$, $\chi_-=0$ to specify $\nematic=\xhat$
in (\ref{eq:zetaP}). For (\ref{eq:p-general}) we have 
\begin{eqnarray}
  \label{eq:p-general-F}
& \expF = F\cos\alpha\sin\beta \xhat + F\sin \alpha\sin\beta \yhat +
F\cos\beta \zhat,\\
  \label{eq:p-general-d}
    & \nematic = (\cos\alpha \cos\beta \cos\gamma
       - \sin\alpha \sin\gamma)\xhat \\\nonumber
       & + (\sin\alpha \cos\beta \cos\gamma +
       \cos\alpha \sin\gamma)\yhat
       -\sin\beta\cos\gamma \zhat.
\end{eqnarray}
Equation~(\ref{eq:p-general}) represents a spinor wave function that
takes values between the FM and 
the polar phases while allowing the spatial variation of the
orientation of the triad and the BEC phase, as determined by
$(\phi,\alpha,\beta,\gamma)$. In the absence of the Zeeman shifts, it
gives all the degenerate states. The Zeeman energy contribution can
partially lift this degeneracy, but as we will consider non-uniform
defect states in a rotating trap, (\ref{eq:p-general}) provides the
most suitable starting point for constructing the initial states for
the energy minimization. 

We can now construct specific defect configurations, that
connect FM and polar defects by making appropriate choices for
$(\phi,\alpha,\beta,\gamma)$. All the basic defect connections that we
have engineered 
are presented in table~\ref{Table:p-Analytics}. The elementary defect
and textures of the spin-1 system that act as building blocks are
briefly summarized in \ref{sec:appendix}. 
Here we give an explicit
discussion of some representative examples. The procedure for
constructing the vortex connections is to
first identify the essential characteristics of the limiting defect
states and then the necessary parameter choices
in~(\ref{eq:p-general}).

\emph{Singly quantized vortex
penetrating the interface:} The phase vortex, formed by a $2\pi$ winding of
the condensate phase alone, corresponds
to a singular, singly quantized vortex in both FM and polar limits
(see~\ref{sec:appendix}). 
Hence, we may continuously connect the two across the interface formed
as $p$ is varied by choosing
$\phi=\varphi$, where $\varphi$ is the azimuthal angle in polar
coordinates, and keeping the Euler angles $\alpha$, $\beta$ and
$\gamma$ constant in~(\ref{eq:p-general}). Making the simplifying assumption
$\alpha=\gamma=0$, we then have
\begin{equation}
\label{eq:p-sing}
\zeta=\frac{e^{i\varphi}}{2}
      \threevec{\sqrt{2} \left(\sin^2\frac{\beta}{2}P_-
		-\cos^2\frac{\beta}{2}P_+\right)}
	       {-\sin\beta\left(P_- + P_+\right)}
	       {\sqrt{2}\left(\cos^2\frac{\beta}{2}P_-
		 - \sin^2\frac{\beta}{2}P_+\right)}.
\end{equation}
Note that this solution is deceptively
simple: a singly quantized vortex represents entirely
different objects (due to the different topology) in the two phases.
More complicated vortex states can be constructed by more
elaborate choices.

\emph{Singly quantized polar vortex to FM coreless vortex:}
The latter is characterized by a $2\pi$ winding of the
condensate phase, together with a simultaneous spin rotation
represented by a $2\pi$ winding of $\alpha$, as described
by~(\ref{eq:cl}). Hence we choose
$\phi=\alpha=\varphi$ ($\gamma=0$), giving
\begin{equation}
\label{eq:p-cl-to-sing}
\zeta=\frac{1}{2}
      \threevec{\sqrt{2} \left(\sin^2\frac{\beta(\rho)}{2}P_-
		- \cos^2\frac{\beta(\rho)}{2}P_+\right)}
	       {-e^{i\varphi}\sin\beta(\rho)\left(P_- + P_+\right)}
	       {\sqrt{2}e^{i 2\varphi}\left(\cos^2\frac{\beta(\rho)}{2}P_-
		 - \sin^2\frac{\beta(\rho)}{2}P_+\right)},
\end{equation}
where we also require $\beta(\rho)$ to increase
monotonically with the radial distance $\rho$, from $\beta=0$ on the $z$
axis, to form the characteristic fountain-like spin texture.
In the polar limit, a $2\pi$ condensate-phase winding represents a singly
quantized vortex.
According to~(\ref{eq:p012}), the
remaining $2\pi$ winding in $\alpha$ only associates a rotation of the
$\nematic$-vector with the singly quantized vortex. As $p$ varies
between $0\le \abs{p}\le c_2n$, this singly quantized
polar vortex connects across the
interface to the coreless vortex in the FM limit.

\emph{Termination of a singular FM vortex as a point defect} on the
interface.  A point defect in the polar phase
corresponds to a radial hedgehog of the $\nematic$ axis, analogous to
the 't~Hooft-Polyakov
monopole~\cite{t-hooft_npb_1974,polyakov_jetplett_1974}. The simplest
example is given
in~(\ref{eq:nematic-monopole}).  In the polar limit
of~(\ref{eq:p-general}), we form the point defect by a $2\pi$ winding
in $\alpha$
together with $\beta=\theta-\pi/2$, where $\theta$ is the polar angle
in spherical coordinates.  In
the FM limit the same choices correspond to a singular spin texture
similar to~(\ref{eq:sv}), exhibiting a
radial disgyration around the singular line.  We can thus construct a
singular FM vortex that terminates as the upper half of a polar point
defect by choosing $\alpha=\varphi$ and $\beta=\theta-\pi/2$.

\emph{Half-quantum vortex to singular FM vortex:} The defining feature of
a polar half-quantum vortex~(\ref{eq:hq}) is a $\pi$ winding of the
condensate phase $\phi$, together with a simultaneous
$\nematic\to-\nematic$ winding of 
the nematic axis to keep the order parameter single-valued.  However,
no similar construction is possible in the FM phase. Therefore the winding
of the condensate phase must combine with the spin rotation
represented by the third Euler angle $\gamma$ to make the combined
$\phi^\prime=\phi-\gamma$ in the FM limit [cf.~(\ref{eq:ferro})] wind by
a multiple of $2\pi$.
The combination $\phi=-\gamma=\varphi/2$
($\alpha=0$) connects the half-quantum vortex to a singly quantized
vortex defined by $\phi^\prime=\varphi$ in the FM limit.

\emph{Terminating half-quantum vortex:} If we instead let
$\gamma=\phi=\varphi/2$, so that these enter the spinor with the same
sign, the polar limit of~(\ref{eq:p-general}) remains a half-quantum
vortex, with the rotation
of $\nematic$ being in the opposite sense. However, in the FM limit,
$\phi$ and $\gamma$ now cancel, $\phi^\prime=0$, and the order
parameter represents a vortex-free state.  The half-quantum vortex in
the polar part thus terminates at the interface.
\begin{table}[tb]
\caption[Analytically constructed interface-crossing defect
  configurations in polar
and FM limits]{
  Interface-crossing defects in the polar interaction regime ($c_2>0$)
  are constructed from~(\ref{eq:p-general}) by different choices for
  $\phi$, $\alpha$, and $\gamma$ (given as multiples of the azimuthal
  angle $\varphi$). For states with non-constant $\beta$, its
  functional form is given in the table, where $\beta(\rho)$ denotes a
  monotonically increasing function of the radial distance only (see
  text for details). The two solutions with a Dirac monopole in the FM
  limit differ by aligning the doubly quantized Dirac string with the
  positive and negative $z$ axis, respectively [cf.~(\ref{eq:dirac})].}
\begin{indented}
\item[]\begin{tabular}{@{}llllll}
\br
FM limit & Polar limit & $\phi/\varphi$ & $\alpha/\varphi$ &
$\gamma/\varphi$ & $\beta$\\
\mr
Vortex free & Half-quantum vortex & $1/2$ & $0$ & $1/2$ & const.\\
Vortex free & Singly quantized vortex & $1$ & $0$ & $1$ & const.\\
Coreless vortex & Half-quantum vortex & $1/2$ & $1$ & $-1/2$ & $\beta(\rho)$\\
Coreless vortex & Singly quantized vortex & $1$ & $1$ & $0$ & $\beta(\rho)$\\
Coreless vortex & Nematic coreless vortex & $0$ & $1$ & $-1$ &
$\beta(\rho)$\\
Singular vortex & Nematic coreless vortex & $0$ & $1$ & $0$ &
$\beta(\rho)$\\
Singular vortex & Half-quantum vortex & $1/2$ & $1$ & $1/2$ &
const.\\
Singular vortex & Half-quantum vortex & $1/2$ & $0$ & $-1/2$ & const.\\
Singular vortex & Singly quantized vortex & $1$ & $0$ & $0$ & const.\\
Singular vortex & Point defect & $0$ & $1$ & $0$ &
$\beta=\theta-\frac{\pi}{2}$\\
Dirac Monopole ($z_+$) & Singly quantized vortex & $-1$ & $1$ & $0$ & $\beta=\theta$\\
Dirac Monopole ($z_-$) & Singly quantized vortex & $-1$ & $-1$ & $0$ & $\beta=\theta$\\
\br
\end{tabular}
\end{indented}
\label{Table:p-Analytics}
\end{table}

As shown in table~\ref{Table:p-Analytics}, we also find solutions of a
terminating singly quantized polar vortex, a half-quantum vortex
connecting to a coreless vortex, a nematic coreless
vortex~(\ref{eq:nematic-coreless}) connecting either to a coreless or
a singular vortex, and a Dirac monopole~(\ref{eq:dirac}) continuously
perforating the interface to a singly quantized polar vortex. Note
that the vortex line (Dirac string) attached to the Dirac monopole may
be formed in two ways: It can be included in the FM phase in such a
way that the Dirac monopole joins the polar vortex to the FM vortex
that forms the Dirac string. Alternatively, the polar vortex itself
can act as a Dirac string, so that no other vortices need to be
coupled to the monopole. In the latter case, the polar vortex
terminates on the interface to a point defect.

\section{Interface by quadratic Zeeman shifts for $c_2<0$}
\label{sec:q-analytics}

In the FM interaction regime ($c_2<0$), we proceed as in the polar case,
but now transform the spinor wave function~(\ref{eq:qsol}) by applying
a BEC phase 
$\phi$ and rotations ($\alpha,\beta,\gamma$) of the spinor to obtain
\begin{eqnarray}
  \label{eq:q-general}
  \zeta_\pm = &\frac{e^{i\phi}}{2\sqrt{2}}\left[
    e^{\mp i(\alpha+\gamma)}(\pm p+q)Q_+\cos^2\frac{\beta}{2}
    + e^{\mp i(\alpha-\gamma)}(\mp p+q)Q_+\sin^2\frac{\beta}{2}\right.\\\nonumber
    &\left.\mp e^{\mp i\alpha}\sqrt{q^2-p^2}Q_-\sin\beta \right],\\
  \zeta_0 = &\frac{e^{i\phi}}{4}\left\{
    2\sqrt{q^2-p^2}Q_-\cos\beta
    + \left[e^{-i\gamma}(p+q)+e^{i\gamma}(p-q)\right]Q_+\sin\beta
    \right\},
\end{eqnarray}
where
\begin{equation}
  Q_\pm = \sqrt{\frac{-p^2 \pm q^2 + 2c_2nq}{2c_2nq^3}}.
\end{equation}
Equation~(\ref{eq:q-spin0}) gives the local spin direction
$\mathbf{F}^0$ before the spin rotation, in terms of which the
general spin texture can be expressed as
\begin{samepage}
\begin{eqnarray}
  \label{eq:q-spin-general}
  \expF =
  &\left[(\cos\alpha\cos\gamma\cos\beta-\sin\alpha\sin\gamma)F^0_x
    + \cos\alpha\sin\beta F^0_z\right]\xhat \\\nonumber
  &+ \left[(\cos\alpha\sin\gamma\cos\beta+\sin\alpha\cos\gamma)F^0_x
    + \sin\alpha\sin\beta F^0_z\right]\yhat \\
  &+ \left[-\cos\alpha\sin\beta F^0_x+\cos\beta F^0_z\right]\zhat. \nonumber
\end{eqnarray}
\end{samepage}
A corresponding expression for $\nematic$ may be derived by rotating
the $\nematic$-vector of~(\ref{eq:qsol}).
We can now make particular choices for $\alpha$, $\beta$,
$\gamma$ and $\phi$ in order to construct specific
defect states.  The basic interface-crossing defect configurations are
presented in table~\ref{Table:q-Analytics}. The derivation is very
similar to the polar 
case and we only provide a brief example and highlight the differences
in the case of half-quantum vortices.

\emph{FM coreless to singly quantized polar vortex:} To form the
coreless vortex~(\ref{eq:cl}) in the FM phase, we require
$\alpha=\varphi$ together with a winding 
$\phi^\prime=\phi-\gamma=\varphi$, as in the $c_2>0$ case.  Note,
however, that this equivalence between rotations of
$\phi$ and $\gamma$ holds only in the purely FM limit, and assigning
the $2\pi$ winding to $\phi$ or $\gamma$ leads to different vortex
states in the polar limit (see
table~\ref{Table:q-Analytics}).  In the former case,
with $\gamma=0$, we have
\begin{eqnarray}
  \label{eq:q-cl-to-sing}
  \zeta_+ = &\frac{1}{2\sqrt{2}}\left[
    (p+q)Q_+\cos^2\frac{\beta(\rho)}{2}
    + (-p+q)Q_+\sin^2\frac{\beta(\rho)}{2}\right.\\\nonumber
    &\left. - \sqrt{q^2-p^2}Q_-\sin\beta(\rho) \right],\\\nonumber
  \zeta_0 = &\frac{e^{i\phi}}{2}\left[
    \sqrt{q^2-p^2}Q_-\cos\beta(\rho)
    + pQ_+\sin\beta(\rho) \right],\\\nonumber
  \zeta_- = &\frac{e^{2i\varphi}}{2\sqrt{2}}\left[
    (-p+q)Q_+\cos^2\frac{\beta(\rho)}{2}
    + (p+q)Q_+\sin^2\frac{\beta(\rho)}{2}\right.\\\nonumber
    &\left. + \sqrt{q^2-p^2}Q_-\sin\beta(\rho) \right].
\end{eqnarray}
Similarly to~(\ref{eq:p-cl-to-sing}), monotonically increasing
$\beta(\rho)$ yields the required fountain-like texture in the FM
limit. In the polar limit,
the winding of the condensate phase implies that
(\ref{eq:q-cl-to-sing}) reduces to~(\ref{eq:p012}), representing a
singly quantized vortex, with which a $2\pi$ winding of $\nematic$
is associated.

\emph{Half-quantum vortices:} The polar half-quantum vortices may
connect across 
the interface to coreless or singular vortices, or
terminate at the interface, as in the polar
interaction regime in section~\ref{sec:p-analytics}.  The analytic
construction of these states from~(\ref{eq:q-general}) in the FM
interaction regime is less straightforward, as the dependence on
$\gamma$ in this case vanishes in the polar limit.
The required $\pi$ winding of $\nematic$
must therefore instead be specified as $\beta = \varphi/2$, and
$\beta$ must then vary differently on the opposite sides of the
interface, such that the wave function 
remains single-valued in the FM limit.  These states will not be
considered further here.
\begin{table}[tb]
\caption{Interface-crossing defect configurations in the FM
  interaction regime ($c_2<0$) 
  are constructed from~(\ref{eq:q-general}) by different choices for
  $\phi$, $\alpha$, and $\gamma$ (given as multiples of the azimuthal
  angle $\varphi$, except for $\gamma=\pi$). For states with
  non-constant $\beta$, its
  functional form is given in the table, where $\beta(\rho)$ denotes a
  monotonically increasing function of the radial distance only (see
  text for details). The two solutions with a Dirac monopole in the FM
  limit differ by aligning the doubly quantized Dirac string with the
  positive and negative $z$ axis, respectively
  [cf.~(\ref{eq:dirac})]. Solutions involving half-quantum vortices
  are omitted since they cannot be straightforwardly constructed (see text).}
\begin{indented}
\item[]\begin{tabular}[tb]{@{}llllll}
\br
FM limit & Polar limit & $\phi/\varphi$ & $\alpha/\varphi$ & $\gamma/\varphi$ & $\beta$\\
\mr
Vortex free & Singly quantized vortex & $1$ & $0$ & $1$ & $\mathrm{const.}$\\
Coreless vortex & Singly quantized vortex & $1$ & $1$ & $0$ & $\beta(\rho)$\\
Coreless vortex & Nematic coreless vortex & $0$ & $1$ & $-1$ & $\beta(\rho)$\\
Singular vortex & Singly quantized vortex & $1$ & $0$ & $0$ & $\mathrm{const.}$\\
Singular vortex & Point defect & $0$ & $1$ & $\gamma=\pi$ & $\beta=\theta$ \\
Singular vortex & Nematic coreless vortex & $0$ & $1$ & $0$ & $\beta(\rho)$\\
Dirac monopole ($z_+$) & Singly quantized vortex & $-1$ & $1$ & $0$ & $\beta=\theta$\\
Dirac monopole ($z_-$) & Singly quantized vortex & $-1$ & $-1$ & $0$ & $\beta=\theta$\\
\br
\end{tabular}
\end{indented}
\label{Table:q-Analytics}
\end{table}

\section{Preparation of vortex states}
\label{sec:imprinting}

Several techniques have been proposed for controlled preparation of
vortex states in BECs.
These include transfer of angular momentum using Laguerre-Gaussian
laser beams~\cite{bolda_pla_1998,marzlin_prl_1997,dutton_prl_2004}, combining
mechanical rotation with coupling to an electromagnetic
field~\cite{williams_nature_1999}, and rotation of the atomic spins by
inverting a magnetic axial bias field~\cite{isoshima_pra_2000}.
Experimental implementations have demonstrated
phase-imprinting of both singly and doubly quantized vortex
lines~\cite{matthews_prl_1999,leanhardt_prl_2002,shin_prl_2004,andersen_prl_2006},
and in spinor BECs also preparation of non-singular
textures~\cite{leanhardt_prl_2003,leslie_prl_2009,choi_prl_2012,choi_njp_2012}.
These
existing techniques could be used also to prepare defect states
when an interface established by a non-uniform Zeeman shift is present.
However, the relation between the analytically constructed defect
solutions and the phase-imprinted states is different in the two
interaction regimes ($c_2 \gtrless 0$).

In the polar interaction regime, the solutions of
section~\ref{sec:p-analytics}  straightforwardly correspond to spin
rotations of (\ref{eq:zetaP}).  Together with the condensate phase
these result in singly or doubly quantized vortex lines in the
individual spinor components, which may be directly phase imprinted using the
existing techniques.  For example, the interface-penetrating singly
quantized vortex corresponds to a singly quantized vortex line in each
of the spinor components.  To connect a singly quantized polar vortex
to a FM coreless vortex instead, vortex
lines with phase winding of $2\pi$ and $4\pi$ respectively are
imprinted in the $\zeta_{0,-}$ components [cf.~(\ref{eq:p-cl-to-sing})].

The preparation of vortex states in the FM interaction regime is less
straightforward.  Due to the spin rotation implicit in the
interpolating ground-state solution (\ref{eq:qsol}), the analytically
constructed defect solutions cannot easily be phase imprinted
directly.  However, phase-imprintable defect wave functions
representing the same defect states can be constructed by considering
a target defect state in the FM or polar
limit~\cite{borgh_prl_2012,borgh_pra_2013}.

Consider, e.g., the singly quantized FM vortex, constructed as a
$2\pi$ winding of the condensate phase. For suitably chosen
parameters, changing the sign of either of $\zeta_\pm$ causes the
vortex wave function to switch from $\absF=1$ to $\absF=0$, such that
it instead represents a singly quantized polar vortex. We can thus join
the singly quantized vortices of the FM and polar phases by switching
the sign of, e.g., $\zeta_-$ at the position of the interface to form
\begin{equation}
  \label{eq:1qs}
  \zeta^{\rm 1\leftrightarrow s} =
  \frac{e^{i\varphi}}{\sqrt{2}}
  \threevec{\sqrt{2}e^{-i\alpha}\cos^2\frac{\beta}{2}}
           {\sin\beta}
           {\mp\sqrt{2}e^{i\alpha}\sin^2\frac{\beta}{2}},
\end{equation}
using the negative sign in the polar part of the condensate, and
correspondingly the positive sign in the FM part.  Note that the
change of sign exactly yields a polar wave function only for
$\beta=\pi/2$.  However, also for any other $\sin(\beta)\neq0$, the
spinor wave function exhibits the spinor-component vortex lines
required for the singly
quantized vortex and quickly relaxes to the polar phase.
Physically, the sign change in $\zeta_-$ corresponds to a dark soliton
plane (a phase kink) where the density in that particular spinor
component vanishes.  However,
the density in the other two
spinor components does not simultaneously vanish at the position of
the soliton plane, and hence the full spinor wave function remains
non-vanishing and continuous.

Approximate wave functions corresponding to other defect states may be
constructed analogously.  For example, when a singly
quantized polar vortex is associated with a simultaneous
rotation of the nematic axis, the spinor components exhibit the same
vortex structure as
the coreless FM vortex~(\ref{eq:cl}). Hence by again
inserting a soliton plane in $\zeta_-$, we obtain the interface spinor
\begin{equation}
  \label{eq:s-cl}
  \zeta^{{\rm 1}\leftrightarrow{\rm cl}} =
  \frac{1}{\sqrt{2}}\threevec{-\sin\beta}
                             {\sqrt{2}e^{i\varphi}\cos\beta}
                             {\pm e^{2i\varphi}\sin\beta},
\end{equation}
with the positive sign in the polar phase. With the negative sign, the
wave function approximates the coreless vortex on the FM side, and
quickly relaxes to $\absF=1$ and forms the characteristic
fountain-like spin texture.

The construction is not limited to the connection of line defects across
the interface.  Also wave functions representing vortices terminating
as point defects on the interface can be engineered.  For example,
the polar point defect~(\ref{eq:nematic-monopole}) is formed by
overlapping vortex lines
of opposite winding in $\zeta_\pm$ together with a soliton plane in
$\zeta_0$. The point defect is placed on the interface by introducing a
soliton plane also in $\zeta_+$,
\begin{equation}
\label{eq:sv-pm}
  \zeta^{\rm sv\leftrightarrow pm} =
  \frac{1}{\sqrt{2}}\threevec{\mp e^{-i\varphi}\sin\theta}
                             {\sqrt{2}\cos\theta}
	                     {e^{i\varphi}\sin\theta}\,,
\end{equation}
such that the positive sign yields a wave function where the
overlapping vortex lines approximate the singular vortex~(\ref{eq:sv})
on the FM side. On the polar side, the radial hedgehog
$\nematic=\rhat$ in the nematic axis is retained.

These examples demonstrate that approximations to interface-crossing
defect states may very generally be constructed from elementary
building blocks of singly 
and doubly quantized vortex lines in the individual spinor components,
together with a dark soliton plane (also phase-imprinted in
experiments~\cite{burger_prl_1999,denschlag_science_2000})  at the
position of the interface. 
Engineering vortex connections consisting of half-quantum vortices on
the polar side is more involved. 
The preparation is complicated by the fact
that there are no vortex solutions exhibiting $\pi$ winding of the
Euler angles that parametrize the FM order parameter.  This implies
that the construction will necessitate phase-imprinting of a vortex
line that terminates at a soliton plane in one of the spinor components.
Considering the connection of a half-quantum vortex~(\ref{eq:hq}) to a
coreless vortex~(\ref{eq:cl}), we may then imagine proceeding as
follows: By introducing a soliton plane in $\zeta_+$ in~(\ref{eq:cl}),
we again construct an interface spinor.  However, on the
polar side of the interface, we can now let a vortex line in $\zeta_+$
terminate on the soliton plane.  As a final step, we may use an
optical shift to deplete the $\zeta_0$ component in the polar
part. The coreless vortex~(\ref{eq:cl}) then remains in the FM part of
the cloud, while the spinor on the polar side of the interface
approximates
\begin{equation}
\zeta =
  \frac{1}{\sqrt{2}}\threevec{-e^{-i\varphi}}
                          {0}
                          {e^{2i\varphi}}
  = \frac{e^{i\varphi/2}}{\sqrt{2}}\threevec{-e^{-3i\varphi/2}}
                          {0}
                          {e^{3i\varphi/2}},
\end{equation}
which represents a half-quantum vortex where $\nematic$ exhibits a
$3\pi$ winding into $-\nematic$ as the vortex line is encircled. The
continuity of the spinor wave function across the interface can be
ensured by the $\zeta_-$ component, which exhibits only a doubly
quantized vortex line and no soliton plane, and therefore does not
vanish simultaneously across the entire interface.

\section{Energetic stability and defect core structures}

By the analytical constructions, we have demonstrated the existence of
continuous wave functions 
representing topologically allowed interface-perforating defect
connections (in tables~\ref{Table:p-Analytics}
and~\ref{Table:q-Analytics}).  To determine their energetic
stability, and the corresponding stable core structures, we minimize
the energy of each defect state by integrating 
the coupled Gross-Pitaevskii equations~(\ref{eq:gpe}) 
in imaginary time, in the frame rotating with frequency $\Omega$.  In
experiment, the condensate is 
trapped by a harmonic potential, which we here take to be axially
symmetric and slightly elongated along the $z$ direction:
\begin{equation}
V(\rr)=\frac{m\omega^2}{2}\left(x^2+y^2+\frac{1}{4}z^2\right)\,,
\end{equation}
We take the spin-independent nonlinearity to be $Nc_0=10^4\hbar\omega
l_\perp^3$, where $l_\perp=\sqrt{\hbar/m\omega}$ is the oscillator
length in the transverse direction. We
consider the experimentally relevant cases
$c_0/c_2=28$, corresponding to $^{23}$Na~\cite{knoop_pra_2011}, and
$c_0/c_2=-216$, corresponding to $^{87}$Rb~\cite{van-kempen_prl_2002},
in the polar and FM regimes, respectively.

We can estimate the energy shifts required to establish the interface
from~(\ref{eq:zetaP}) and (\ref{eq:qsol}).  The gradients in $p$ and
$q$ are then determined by the width of the interface region.  In our
numerics we have studied large widths of up to $10 l_\perp$ and find that
the qualitative features of the defect states remain unchanged. 
Since the width can be varied from  large values down to the healing length
scale, the possible values of the field gradient may cover a very large
range of values. The experimentally most promising
method to induce the 
energy shifts themselves is by using electromagnetic dressing fields, as
demonstrated for the quadratic
shift~\cite{gerbier_pra_2006}. Potentially, similar methods could be 
used to manipulate also the linear shift, which may prove
experimentally easier than using a static magnetic field.

For the case of FM interactions, we take the interface to be
established by varying $q$ at constant $p$ 
(cf.\ figure~\ref{fig:q-validity}). We consider the example of
$^{87}$Rb and approximate the density profile by the Thomas-Fermi
solution. Then 
for very small $p\sim 10^{-3}\hbar\omega$, the necessary difference in $q$
is $\sim 0.15\hbar\omega$. 
For larger $p$ the required change in $q$ is smaller (while $q$
itself is larger).  For the recent experiment~\cite{guzman_pra_2011}
the induced level shift is given in terms of the Rabi
frequency $\Omega_\mathrm{R}$ and detuning $\delta$ as
$q=-\hbar\Omega_\mathrm{R}^2/4\delta$, with 
$|\delta|=2\pi\times40$~kHz. As an example, we may consider a trap frequency
$\omega = 2\pi\times50$~Hz. We then find 
$\Omega_\mathrm{R}^\mathrm{polar}-\Omega_\mathrm{R}^\mathrm{FM} 
\simeq 6$~kHz.  

In the case of polar interactions~(\ref{eq:zetaP}), $\absF$ 
depends only on the linear Zeeman shift $p$, and reaches the FM phase
for $p \geq c_2n$.  Here we consider the example of $^{23}$Na.
Then the necessary shift in $p$ is $\sim 0.5\hbar\omega$ at the
maximum value of the Thomas-Fermi density.
Considering again the example $\omega = 2\pi\times50$~Hz,
this corresponds to a field gradient on the order of $1.4$--$14$~G/m,
for the corresponding variation $1$--$10
l_\perp$ of the interface width, if the shift is induced by a weak
static magnetic field.

\subsection{Polar interactions}

When the spin-dependent interaction favours the polar phase, the interface is
created by a spatially varying 
$0 \leq \abs{p} \lesssim c_2n$, corresponding to the ground-state
solution~(\ref{eq:zetaP}).  We then take the wave functions
constructed in section~\ref{sec:p-analytics} as initial states for the
numerical energy minimization.

Even though singly quantized vortices exhibit similar
winding of the condensate phase in both polar and FM phases, their
energy-minimizing core structures are quite
different~\cite{lovegrove_pra_2012}.  In the polar
BEC, the vortex may split to form an
extended core region in which the wave function is excited out of the
ground-state manifold. It reaches the FM phase on two singular
half-quantum vortex lines.  This lowers the energy by allowing the
core size to be determined by the spin healing length
$\xi_F$, defined in~(\ref{eq:xi}), which is usually 
larger than the size of a density-depleted core, given by the density
healing length $\xi_n$. Also in the
FM phase the core of the singly quantized vortex can avoid the density
depletion.  However, here the defect cannot split.  Instead, filling of
the vortex core happens by local rotation of the spin vector around
the vortex line. The overall structure maintains the
axial symmetry of the vortex core.

\begin{figure}[tb]
  \begin{center}
    \includegraphics[scale=1]{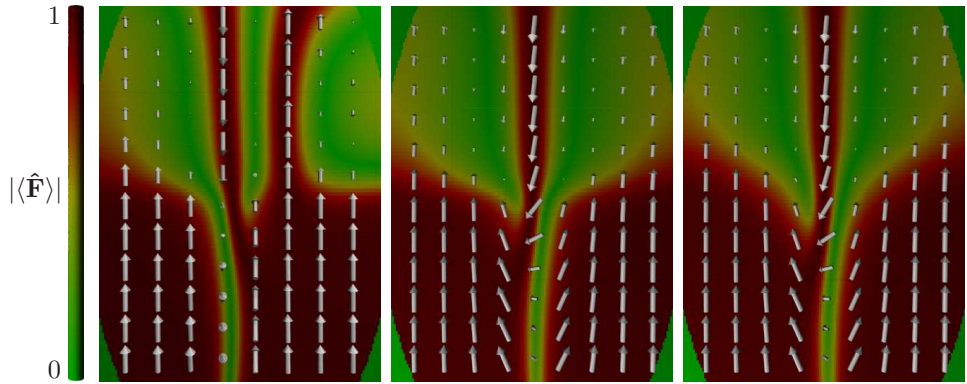}
  \end{center}
  \caption{
    Spin vector (arrows) and spin magnitude (colour gradient) in the
    energetically stable connection of a
    singly quantized (left) or half-quantum (middle and right) polar
    vortex to a singly quantized FM vortex. In the former case the core of
    the singly quantized polar vortex lowers its energy by splitting into
    a pair of half-quantum vortices. The singular FM vortex reduces
    the 
    energy of the core by filling with the polar phase at the singularity.
    Across the interface region, the linear Zeeman shift varies as
    $1.0\times10^{-3}\hbar\omega\leq p\leq 0.6\hbar\omega$, with a
    constant quadratic shift $q=-1.0\times10^{-4}\hbar\omega$.  The linear
    shift varies between the polar and FM limits over a distance $1.0
    l_\perp$ (left and middle) and $4.0 l_\perp$ (right), respectively,
    showing that the qualitative defect structure is insensitive to the
    width of the interface region. The rotation frequency of the system is
    $\Omega=0.22\omega$ for the singly quantized polar vortex
    and $\Omega=0.20\omega$ for the half-quantum vortex.
  }
  \label{fig:p-singular}
\end{figure}

\emph{Interface-crossing singly quantized vortex:}
When part of the condensate is forced into the FM phase by a linear
Zeeman shift, these deformation mechanisms lead to a complex,
energetically stable vortex configuration as the energy of an
interface-crossing singly quantized vortex relaxes
(figure~\ref{fig:p-singular}, left).  The splitting instability leads to the
formation of two vortex lines filled with the FM phase on the polar
side of the interface. In the FM region, the vortex core fills with the polar
phase in order to lower its energy.  The filling of the core is made
possible by a local rotation of the spin vector close to the vortex
line. The resulting spin profile connects smoothly to the spin
vector in the FM cores of the polar vortices at the interface.

\emph{Half-quantum vortex to singular FM vortex:}
A similar penetration of the FM phase through the interface to fill
the singular line in the polar order parameter occurs in the
energetically stable
connection of a polar half-quantum vortex to a singular FM vortex
(figure~\ref{fig:p-singular}, middle and
right). Simultaneously, the singularity in the FM phase
fills with the polar phase in order to minimize its associated
gradient energy.  Consequently, at the perforation of the interface
the two core
structures meet and connect to the ground-state phase on the other side
of the boundary. Figure~\ref{fig:p-singular} also shows that the
qualitative features of the defect connection is not contingent on a
sharply defined interface region, which is a general feature of our
stable defect configurations.

\emph{Singly quantized (or half-quantum) vortex to coreless vortex:}
In the purely FM spin-1 BEC, the singular vortex can be energetically
(meta-)stable, but a lower-energy coreless vortex generally exists for
the same parameters~\cite{lovegrove_pra_2012}.  One might
therefore expect stable states to
exist where a coreless vortex connects across the FM-polar interface
to a singly quantized (or half-quantum) polar vortex.  
However,
the linear Zeeman shift that is employed here to realize the FM phase
when $c_2>0$ makes
the fountain-like spin texture of the coreless vortex energetically
unfavourable. As a consequence, we find all vortex connections involving a
coreless vortex on the FM side to be unstable. 

\begin{figure}[tb]
  \begin{center}
    \includegraphics[scale=1]{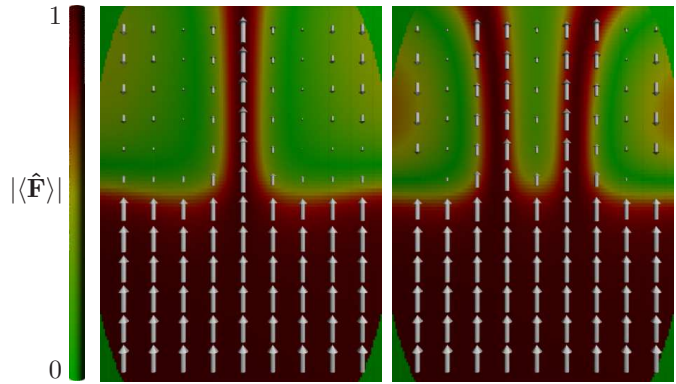}
  \end{center}
  \caption{
    Spin vector (arrows) and spin magnitude (colour gradient) showing
    an energetically stable half-quantum (left) or singly
    quantized (right) polar vortex terminating at the interface between the
    polar and FM parts of the condensate.  Energy relaxation causes the
    singly quantized vortex to split into a pair of half-quantum
    vortices. The linear shift varies over
    $1.0\times10^{-3}\hbar\omega\leq p\leq 0.5\hbar\omega$ across an
    interface of width
    $1.0l_\perp$, with $q=-1.0\times10^{-4}\hbar\omega$.  The system rotates at
    $\Omega=0.22\omega$ for the singly-quantized polar vortex
    and $\Omega=0.2\omega$ for the half-quantum vortex.
  }
  \label{fig:p-terminating}
\end{figure}

\emph{Terminating polar vortices:}
The constructions in section~\ref{sec:p-analytics}
(table~\ref{Table:p-Analytics}) demonstrate that it
is also possible for polar vortices to terminate at the interface.
The terminating half-quantum vortex is energetically stable, and the
relaxed core structure is shown in figure~\ref{fig:p-terminating} (left).
In the vortex-free FM region, the linear
Zeeman energy causes $\expF$ to align with the $z$ axis, and the FM
phase penetrates the interface to fill the singular core of the polar
vortex.  Also a terminating singly quantized vortex results in a
stable defect configuration (figure~\ref{fig:p-terminating}, right).
In this case, however, relaxation of the
energy causes the singly-quantized vortex to split (preserving
topology) into a pair of half-quantum vortices, whose singular cores fill with
the FM phase.

\begin{figure}[tb]
  \begin{center}
    \includegraphics[scale=1]{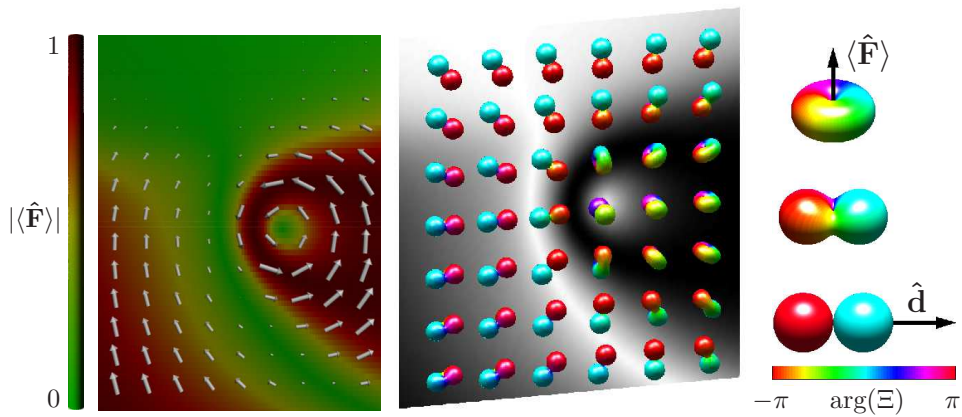}
  \end{center}
  \caption{
    Left: Spin texture (arrows) and magnitude profile (colour gradient)
    for the polar monopole in a polar BEC with interface created by
    varying linear Zeeman splitting.  Right: The order-parameter symmetry
    is shown by mapping $\zeta$ onto the $Y_{1,m}$ spherical
    harmonics~\cite{kawaguchi_physrep_2012}. The function
    $\Xi(\theta,\varphi) = \sum_{m}\zeta_m Y_{1,m}(\theta,\varphi)$ is
    shown on the far right for the FM (top) and polar (bottom) phases and
    for intermediate $\absF$, indicating the $\nematic$-vector. In the
    relaxed defect
    state, $\nematic$ passes through the arch and points radially away
    outside it, preserving the monopole charge.
    Spin magnitude indicated in grey scale.
    Parameters are
    $1.0\times10^{-3}\hbar\omega\leq p\leq 0.3\hbar\omega$, for
    $q=-1.0\times10^{-4}\hbar\omega$ in a
    non-rotating system with interface width $1.0l_\perp$.}
  \label{fig:p-nematic-monopole}
\end{figure}

\emph{A singular FM vortex terminating as a point defect}
exhibits a particularly non-trivial deformation of the defect core as
the energy relaxes.  In order for the core of the point defect to fill
with the FM phase, it deforms into a line defect that
forms a ring-shaped vortex (figure~\ref{fig:p-nematic-monopole})
attached to the interface~\cite{borgh_prl_2012,borgh_pra_2013}.
This is a consequence of the `hairy-ball theorem': if the core of
the point defect were to fill with the FM phase, the spin vector in
the core would have to be everywhere perpendicular to the radial
$\nematic$-vector, which is not possible.
After the deformation, a disclination plane in $\nematic$ may be
identified, such that on any closed loop through the arch formed by
the defect, $\nematic$ winds into $-\nematic$.  Hence, the line defect
is a half-quantum vortex, and the
charge of the point defect is preserved away from the vortex arch.
This phenomenon is
closely related to the similar deformation of a spherically symmetric
point defect into a half-quantum vortex ring---an Alice ring---in the
polar spin-1
BEC~\cite{ruostekoski_monopole_2003}.
The deformation of the point defect into a semi-circular `Alice
arch' on the interface was analyzed also
in~\cite{borgh_prl_2012,borgh_pra_2013}, and our result here
demonstrates that it could be engineered by the Zeeman energy
shifts. Although the defect is stable in the bulk medium, the density
gradient in the trapped condensate 
causes the arch-shaped line defect to be unstable towards drifting out
of the cloud, since a smaller atom density lowers the gradient energy
associated with the core. The defect could be stabilized by reversing
the density gradient with a pinning laser~\cite{ruostekoski_monopole_2003}.

Other defect connections described in table~\ref{Table:p-Analytics}
are found to be energetically unstable. These include, in addition to
FM coreless vortices, also connections involving nematic corelss
vortices and Dirac monopoles.

So far we have considered the energy minimization of initially
prepared vortex configurations. 
In a sufficiently rapidly rotating system, vortices may also nucleate.
In the polar interaction regime, at low enough rotation frequency, we observe
nucleation of a single half-quantum vortex that terminates at the interface.
One might expect this to connect across the interface to a coreless
vortex in the FM phase.  However, this configuration is
energetically less favourable, due to the linear Zeeman-energy cost of
forming the fountain-like coreless spin texture.

\subsection{FM interactions}

We now explore the stability properties and core structures in the FM
interaction regime ($c_2<0$), as for $^{87}$Rb.  Due to the
different ground-state properties of the interpolating solutions, the
interface is now created by a non-uniform quadratic Zeeman shift that
forces the condensate into the polar phase.
Correspondingly we minimize the energy of the defect solutions of
section~\ref{sec:q-analytics} and the corresponding phase-imprinted
configurations of~\ref{sec:imprinting}

\begin{figure}[tb]
  \begin{center}
    \includegraphics[scale=1]{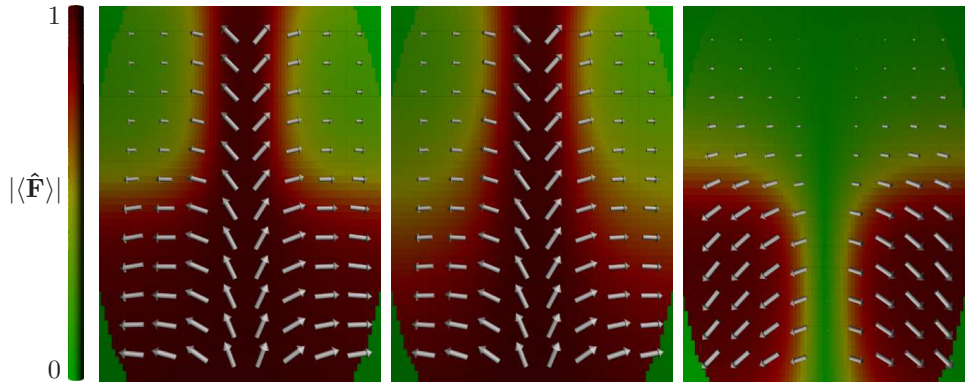}
  \end{center}
  \caption{
    Left and middle: Spin texture (arrows) and spin magnitude (colour
    gradient) of an
    energetically stable coreless vortex connecting to a
    singly quantized polar vortex in a system rotating at
    $\Omega=0.18\omega$.
    The width of the interpolating region
    is $2.0 l_\perp$ (left) and $10.0 l_\perp$
    (middle), over which the quadratic energy shift varies as
    $p\leq q\leq 0.198\hbar\omega$, for
    $p=9.9\times10^{-4}\hbar\omega$. The defect structure is qualitatively
    insensitive to the width of the interface region.
    Right: A singular FM vortex that terminates at the interface in the
    system rotating at $\Omega=0.16\omega$ (remaining parameters same as in
    the left panel). Also the
    connection of a singular FM vortex to a nematic coreless vortex
    relaxes to the state shown as the quadratic Zeeman energy causes
    the fountain texture of the nematic axis to be lost in the polar
    region.}
  \label{fig:q-coreless}
\end{figure}

\emph{Coreless vortex to polar singly quantized vortex:}
Contrary to the polar interaction regime, we now do find an
energetically stable connection of a coreless vortex
on the FM side of the interface to a polar singly quantized vortex
(figure~\ref{fig:q-coreless}), as the energy
of~the initial state (\ref{eq:q-cl-to-sing}) relaxes.
One might again expect the singly
quantized vortex in the polar phase to split into a pair of
half-quantum vortices in order to lower the energy of the core.
However, the splitting is energetically unfavourable due to the
positive quadratic Zeeman shift needed to realize the polar phase in
the BEC with FM interactions, which seeks to align $\nematic$ with the
$z$ axis, resulting in an effective two-component regime.
Accordingly, the stable configuration
exhibits an axially symmetric single core with the atoms reaching the
FM phase at the line singularity. 
The spin texture of the coreless vortex
connects smoothly to a similar spin texture inside the core of the
polar vortex, and is qualitatively independent of the width of the
interface region, as shown in figure~\ref{fig:q-coreless}.

Previous studies of singly quantized polar vortices have shown that
stable structures in the absence of the interface would favour core
structures where the vortex line is split into a pair of half-quantum
vortices~\cite{lovegrove_pra_2012}. Here the existence of both split
and unsplit cores, with entirely different symmetries, as stable
vortex cores is reminiscent of the 
vortex core structures of superfluid liquid
$^3$He~\cite{salomaa_rmp_1987}. In superfluid liquid $^3$He, the core
of a singular $B$-phase vortex may retain a non-vanishing superfluid
density by filling with the $A$ phase. This may appear as an axially
symmetric core~\cite{salomaa_prl_1983} at high pressure or with a
broken axial symmetry~\cite{salomaa_prl_1986,thuneberg_prl_1986}, as
experimentally observed in~\cite{kondo_prl_1991}. 

\emph{Interface-crossing singly quantized vortex:}
The connection of a singly quantized polar vortex in a
uniform $\nematic$ texture to a singular FM vortex is not
energetically stable, again in contrast to the case for polar interactions.
Energy relaxation of the vortex formed as a $2\pi$ winding of the
condensate phase everywhere
causes the initially uniform spin texture in the FM region to deform
locally around the
singular vortex line, allowing the condensate to avoid the density
depletion~\cite{lovegrove_pra_2012}. The singular vortex can then
leave the cloud, nucleating a coreless vortex in the
process. Correspondingly, the singly quantized vortex in the polar
part picks up a winding of $\nematic$, and the initial defect state
decays to the connection of a coreless vortex to a singly quantized
polar vortex, similar to figure~\ref{fig:q-coreless}.

\emph{Terminating singular FM vortex:} 
A singular FM vortex may also be written as a winding of $\alpha$ alone
(for some $\beta$),
in which case it can terminate at the interface.
The configuration relaxes to an energetically stable
vortex state whose spin texture is shown in
figure~\ref{fig:q-coreless} (right). The polar phase then penetrates the
interface to fill the core of the FM vortex, allowing the core to
expand and lower its energy.  By including a winding of $\beta$, the
initial defect state may also represent a nematic coreless
vortex~(\ref{eq:nematic-coreless}) on the polar side (see
table~\ref{Table:q-Analytics}).  This is,
however, not stable, as the fountain texture in $\nematic$ unwinds due
to the quadratic Zeeman shift, resulting again in a terminating FM vortex. 

\begin{figure}[tb]
  \begin{center}
    \includegraphics[scale=1]{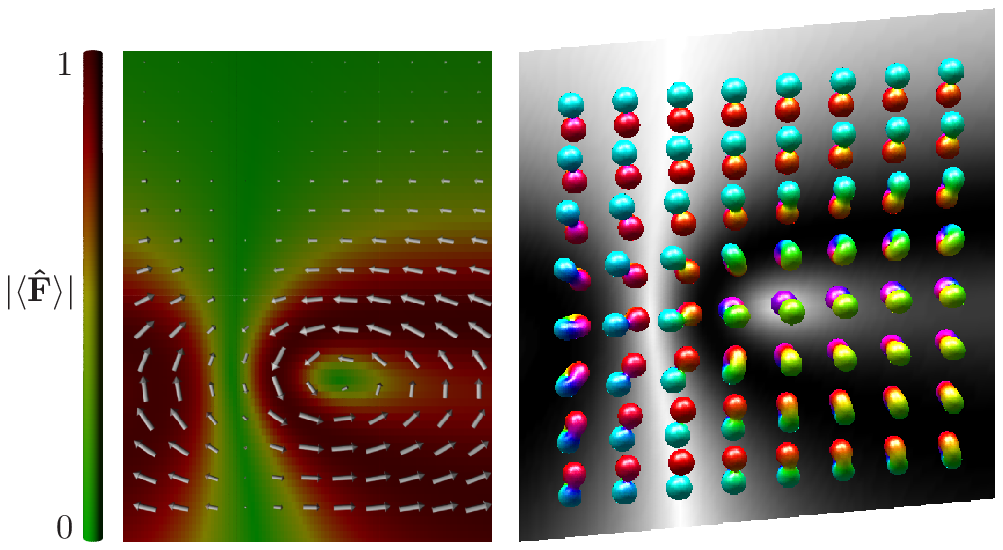}
  \end{center}
  \caption{
    Left: Spin texture $\expF$ (arrows) and magnitude $\absF$ (colour
    gradient) of a
    singular FM vortex connecting to a nematic monopole where the interface
    is induced by quadratic Zeeman splitting. The monopole
    deforms into a line defect as the energy relaxes.
    Right: Mapping of $\zeta$ onto the $Y_{1,m}$ spherical harmonics (see also
    figure~\ref{fig:p-nematic-monopole}), indicating the nematic axis
    $\nematic$ preserving the charge of the monopole. (Away from the
    interface region, $\nematic$ tends towards $\pm\zhat$ due to the
    quadratic Zeeman energy.) Spin magnitude indicated in grey scale.
    Parameters are
    $p\leq q\leq 0.198\hbar\omega$, for $p=9.9\times10^{-4}\hbar\omega$ in a
    non-rotating system with interface width $2.0l_\perp$.}
  \label{fig:q-nematic-monopole}
\end{figure}

\emph{Singular FM vortex terminating as a point defect:}
The relaxed core structure shown in
figure~\ref{fig:q-nematic-monopole} exhibits the deformation of
the point defect into an arch-shaped half-quantum vortex attached to
the interface. This reaches the FM 
phase at the line singularity, and connects to the spin texture of the FM
vortex.  The deformation mechanism is here analogous to that discussed
in the polar case.  Again, the arch-shaped
half-quantum vortex maintains its structure, but is unstable towards
drifting out of the cloud. 

Also in the FM interaction regime, we find that solutions involving a
Dirac monopole are energetically unstable. We further find that
neither the terminating singly 
quantized polar vortex nor the connection of a coreless vortex to a nematic
coreless vortex (see table~\ref{Table:q-Analytics}) are energetically
stable. 

In addition to minimizing the energy of each defect state constructed
in section~\ref{sec:q-analytics}, we also performed
simulations starting from the corresponding experimentally
phase-imprintable defect 
states constructed
from vortex lines and soliton planes in
section~\ref{sec:imprinting}.  In each case, the relaxed defect state
agrees with those resulting from the defect wave functions of
section~\ref{sec:q-analytics}.

As in the polar interaction case, we may also start from a vortex-free
configuration 
and study nucleation of defects as a result of rotation. For weak
rotation we here find nucleation 
of a singly quantized polar vortex that connects to a coreless vortex
in the FM region. 
The stable polar vortex core again preserves the axial symmetry.

\section{Concluding remarks}

In conclusion, we propose that a stable, coherent interface between
topologically dissimilar regions of atomic spinor systems can be
engineered by spatially non-uniform linear or 
quadratic Zeeman shifts, which are commonly manipulated in experiments.
As a particular example
we have shown how an interface can be established between FM and polar
regions of a spin-1 BEC.  We have derived analytic expressions for
states representing continuous defect connections across the
interface, interpolating between FM and polar topology in terms of
either the linear (for $c_2>0$) or quadratic (for $c_2<0$) Zeeman
shifts. We have demonstrated the energetic stability of several
non-trivial interface-crossing defect states.

In the present simulations we did not conserve the longitudinal 
condensate magnetization. In physical systems where $s$-wave
scattering is the dominant relaxation mechanism (compared with, e.g.,
dipole-dipole interactions or collisions with high-temperature atoms),
the magnetization is preserved on experimentally relevant time
scales~\cite{stenger_nature_1998,jacob_pra_2012}. 
Our recent study of coreless
vortices~\cite{lovegrove_prl_2014}, however, indicates that the
conservation of magnetization only plays an important role in
situations where the initial value of the magnetization
differs substantially from the final magnetization values 
obtained in simulations with unconstrained magnetization.
For typical vortex states such conditions can easily be avoided. 

There are several obvious possibilities for
extending our study. The atomic spins generate magnetic dipole
moments. Depending on the atom, these give rise to dipole-dipole
interactions in the condensate, which may influence the structure of
textures and
defects~\cite{santos_spin-3_2006,simula_jpsj_2011}.
Simulations
incorporating these dipole-dipole interactions can be performed by
introducing non-local interactions in the numerical model. 
On the other hand,  
defect formation in annihilation of
colliding interfaces could mimic brane annihilation
scenarios~\cite{dvali_plb_1999,sarangi_plb_2002}. 
Furthermore, defects and textures can be considerably more complex in
spin-2 and spin-3 systems that, for instance, have non-Abelian
vortices~\cite{kobayashi_prl_2009,huhtamaki_pra_2009}. This is likely
to result also in richer interface physics. 
In strongly correlated scenarios,
the atoms may also be confined in optical
lattices 
in such a way that 
interfaces could separate different lattice regions, each
simultaneously exhibiting a
different phase of quantum magnetism.

\ack This work is supported by the Leverhulme Trust and the EPSRC.

\appendix

\section{Elementary vortex solutions}
\label{sec:appendix}
In this appendix we provide for reference a brief overview of the
elementary defect states of the spin-1 BEC in the pure FM and polar
phases. For a more detailed presentation, see,
e.g.,~\cite{kawaguchi_physrep_2012,borgh_pra_2013}.

In the FM phase, all degenerate, physically distinguishable
spinors are related by three-dimensional
spin rotations given by Euler angles $\alpha$, $\beta$ and $\gamma$,
where the third Euler angle is absorbed by the condensate phase in
$\phi^\prime=\phi-\gamma$.  Consequently, an arbitrary FM spinor can
be constructed by applying a spin rotation to a reference spinor
$\zeta=(1,0,0)^T$ to arrive at~(\ref{eq:ferro}), with spin vector
$\expF=\cos\alpha\sin\beta\xhat+\sin\alpha\sin\beta\yhat+\cos\beta\zhat$.
From this general expression, we can construct the non-trivial
representatives of the two classes of line defects supported by the
corresponding $\SO(3)$ ground-state manifold.

The simplest singular vortex corresponds to a $2\pi$ winding of the
condensate phase $\phi^\prime$ in a uniform spin texture.  This phase vortex is
simply described by letting $\phi^\prime=\varphi$, the azimuthal angle,
in~(\ref{eq:ferro}), keeping $\alpha$ and $\beta$ constant.
However,
other, topologically equivalent, singular vortices can be constructed
from the phase vortex by local spin rotations. We may, for example,
rotate the spins into a disgyration corresponding to
$\alpha=\varphi$:
\begin{equation}
  \label{eq:sv}
  \zeta^\mathrm{sv} = \frac{1}{\sqrt{2}}
    \threevec{\sqrt{2}e^{-i\varphi}\cos^2\frac{\beta}{2}}
             {\sin\beta}
             {\sqrt{2}e^{i\varphi}\sin^2\frac{\beta}{2}}.
\end{equation}
When $\beta=0$ the spins align with the $z$ axis, and the singular
spin texture coincides with a phase vortex.  However, for $\beta\neq0$
the spins tilt radially away from the $z$ axis, and at $\beta=\pi/2$
form a radial spin disgyration (spin vortex) that is singular, but
carries no mass circulation.

The fact that mass circulation alone is not quantized in the FM phase
makes it possible for angular momentum to be carried by  non-singular
coreless vortices.  The prototypical coreless vortex is characterized by a
fountain-like spin texture, where the spin aligns with the $z$ axis on
the vortex line, and tilts radially away from it with increasing
radial distance $\rho$, corresponding to a monotonically increasing
$\beta(\rho)$. The wave function is kept non-singular everywhere by
a combined rotation of the spin and the condensate phase,
$\alpha=\phi^\prime=\varphi$, to form
\begin{equation}
  \label{eq:cl}
  \zeta^{\rm cl}(\mathbf{r}) =
  \frac{1}{\sqrt{2}}\threevec{\sqrt{2}\cos^2\frac{\beta(\rho)}{2}}
                          {e^{i\varphi}\sin\beta(\rho)}
                          {\sqrt{2}e^{2i\varphi}\sin^2\frac{\beta(\rho)}{2}}.
\end{equation}
Similarly to the singular vortices, several non-singular vortices are
possible. These are all related to~(\ref{eq:cl}), and to the
vortex-free state, by local spin rotations.

It is further possible to rotate the spins in the coreless vortex to
point everywhere radially away from the origin, resulting in a
terminating, doubly quantized vortex line.  This hedgehog
configuration, $\expF = \rhat$, is
analogous~\cite{savage_dirac_2003,pietila_prl_2009_dirac,ruokokoski_pra_2011}
to the Dirac magnetic monopole~\cite{dirac_pr_1948},
with the doubly quantized vortex line corresponding to the attached Dirac
string. When the Dirac string coincides with the positive $z$ axis, the
corresponding spinor is
\begin{equation}
  \label{eq:dirac}
  \zeta^{\rm D} = \frac{1}{\sqrt{2}}
                  \threevec{\sqrt{2}e^{-2i\varphi}\cos^2\frac{\theta}{2}}
                   {e^{-i\varphi}\sin\theta}
                   {\sqrt{2}\sin^2\frac{\theta}{2}},
\end{equation}
where we have set $\alpha=\varphi$ and $\beta=\theta$, the polar
angle, to form the hedgehog
texture, and chosen $\phi^\prime=-\varphi$.  The Dirac string may
instead be aligned with the negative $z$ axis by instead choosing
$\phi^\prime=\varphi$.

In the polar phase, the order parameter is determined by the
condensate phase and rotations of the nematic axis
$\nematic$, which may be applied to the reference state
$\zeta=(0,1,0)^T$, with $\nematic=\zhat$, to yield
\begin{equation}
  \label{eq:polar}
  \zeta^{\rm p} =
    \frac{e^{i\phi}}{\sqrt{2}}\threevec{-e^{-i\alpha}\sin\beta}
                                       {\sqrt{2}\cos\beta}
                                       {e^{i\alpha}\sin\beta},
\end{equation}
whose equivalence to~(\ref{eq:nematic}) follows from the identification
$\nematic=\cos\alpha\sin\beta\xhat+\sin\alpha\sin\beta\yhat+\cos\beta\zhat$.
Note that the choice of reference state corresponds to the polar limit
of (\ref{eq:qsol}), and the Euler angles in~(\ref{eq:polar})
therefore acquire the same meaning as in the polar limit
of~(\ref{eq:q-general}).  [In the polar limit of~(\ref{eq:p-general}), the
spin rotation is instead applied to the spinor
$\zeta=(-1/\sqrt{2},0,1/\sqrt{2})^T$ with $\nematic=\xhat$, and the
relation between $\nematic$ and the Euler angles is modified accordingly.]

In the polar phase, all circulation-carrying vortices are singular.  The
simplest is again a singly quantized vortex in a uniform
$\nematic$-texture, constructed as $\phi=\varphi$
in~(\ref{eq:polar}).
However, rotations of $\nematic$ do not
contribute to the quantized circulation, and hence a singly quantized
vortex may be accompanied by a winding of $\nematic$ as long as
$\zeta$ remains single valued. For example, the choice
$\phi=\alpha=\varphi$ results in
\begin{equation}
  \label{eq:p012}
  \zeta^{\rm 1^\prime} =
  \frac{1}{\sqrt{2}}\threevec{-\sin\beta}
                             {\sqrt{2}e^{i\varphi}\cos\beta}
                             {e^{2i\varphi}\sin\beta},
\end{equation}
which a singly quantized vortex with a $2\pi$ winding in $\nematic$.

Due to the nematic order
$\zeta(\phi,\nematic)=\zeta(\phi+\pi,-\nematic)$, the single quantum
is not the smallest unit of circulation in the polar phase.  By
combining a $\pi$ winding of the condensate phase with a
$\nematic\to-\nematic$ winding of the nematic axis, one can construct
a vortex carrying half a quantum of circulation. The simplest such
vortex, where $\nematic$ is confined to the $xy$ plane, is represented
by
\begin{equation}
  \label{eq:hq}
  \zeta^{\rm hq} = \frac{e^{i\varphi/2}}{\sqrt{2}}
                   \threevec{-e^{-i\varphi/2}}
                            {0}
                            {e^{i\varphi/2}}
		 = \frac{1}{\sqrt{2}}
                   \threevec{-1}
                            {0}
                            {e^{i\varphi}}.
\end{equation}
In general, a half-quantum vortex may exhibit a more complicated
$\nematic$-field, provided that $\nematic\to-\nematic$ on any closed
loop around the vortex line.

Even though circulation is quantized in the polar phase, it is
possible to form a non-singular nematic coreless
vortex~\cite{lovegrove_prl_2014} that does not carry 
angular momentum. Here  $\nematic$ forms a fountain-like texture
analogous to the FM 
coreless vortex. This structure was recently experimentally phase
imprinted~\cite{choi_prl_2012,choi_njp_2012}. 
  From~(\ref{eq:polar}) it
can be constructed by
choosing $\alpha=\varphi$ combined with $\beta(\rho)$ increasing
monotonically from $\beta(0)=0$ to form
\begin{equation}
  \label{eq:nematic-coreless}
  \zeta^{\rm p} =
    \frac{1}{\sqrt{2}}\threevec{-e^{-i\varphi}\sin\beta(\rho)}
                               {\sqrt{2}\cos\beta(\rho)}
                               {e^{i\varphi}\sin\beta(\rho)}.
\end{equation}

The
polar phase also supports singular point defects (monopoles). The
basic monopole solution is the spherically symmetric $\nematic=\rhat$
texture, which is analogous to the 't~Hooft-Polyakov monopole in
quantum field theory. It is represented by the spinor
\begin{equation}
  \label{eq:nematic-monopole}
  \zeta^{\rm pm} = \frac{1}{\sqrt{2}}\threevec{-e^{-i\varphi}\sin\theta}
                            {\sqrt{2}\cos\theta}
	                    {e^{i\varphi}\sin\theta}.
\end{equation}

\section*{References}

\end{document}